\documentclass[pageno]{jpaper}

\usepackage[normalem]{ulem}
\usepackage{subfig}
\usepackage{comment}
\usepackage{url}

\begin{document}

\title{Towards Offloadable and Migratable Microservices on Disaggregated Architectures: Vision, Challenges, and Research Roadmap}

\author{Xiaoyi Lu and Arjun Kashyap \\
  {\centering\emph{University of California, Merced}}\\
  {\centering{xiaoyi.lu@ucmerced.edu, akashyap5@ucmerced.edu}}}

\date{}
\maketitle

\thispagestyle{empty}

\begin{abstract}

Microservice and serverless computing systems open up massive versatility and opportunity to distributed and datacenter-scale computing. In the meantime, the deployments of modern datacenter resources are moving to disaggregated architectures. With the flourishing growths from both sides, we believe this is high time to write this vision paper to propose a potential research agenda to achieve efficient deployments, management, and executions of next-generation microservices on top of the emerging disaggregated datacenter architectures. In particular, we envision a critical systems research direction of designing and developing offloadable and migratable microservices on disaggregated architectures. With this vision, we have surveyed the recent related work to demonstrate the importance and necessity of researching it. We also outline the fundamental challenges that distributed systems and datacenter-scale computing research may encounter. We further propose a research roadmap to achieve our envisioned objectives in a promising way. Within the roadmap, we identify potential techniques and methods that can be leveraged.

\end{abstract}

\section{Introduction}
\label{sec:intro}

\noindent Microservices and serverless computing are becoming the new paradigm
in cloud computing. The microservice architecture of decoupling monolithic
applications to loosely-coupled microservices is gaining traction in large
datacenters due to the ability to meet strict performance and availability
constraints.  Big cloud providers have already started adopting the
microservice application
model~\cite{twitter-microservice,microservice-workshop,microservice-evo}.
Similarly, many public cloud platforms are offering serverless computing
services like AWS Lambda~\cite{aws-lambda}, Azure
Functions~\cite{azure-function}, and Google Cloud
Functions~\cite{google-cloud-fn} as the platform automatically manages the
infrastructure and server to run these serverless functions.

With the increasing popularity of microservices, current efforts have been put
to study the characteristics, requirements, and implications they have on the
cloud system
stack~\cite{DeathStarBench,ppbench,usuite,workload-microservice,benchmark-microservice}.
Due to the strict service level objectives set (Quality-of-Service
requirements) by customers for throughput and tail latency, microservices put
pressure on cloud architecture ranging from hardware to operating system and
all the way up to application design. Lastly, datacenters have a huge energy
footprint~\cite{e3} and these days the cloud providers equally focus on both
performance and energy-efficiency of the applications to reduce the carbon
footprint.

On the other hand, \textit{Disaggregated Architecture} for modern datacenters
becomes a hot research topic recently.  Disaggregated architecture splits the
existing monolithic datacenter tier into at least two independent resource
pools (i.e., compute pool and storage pool) that communicate over fast
interconnects, such as high-speed Ethernet or InfiniBand. By sharing the
storage pool across diverse datacenter services, disaggregated architecture
consolidates computational and storage resources independently. Thus, it allows
independent optimization and resource scaling, and reduces total cost. The cost
and efficiency benefits of disaggregated architecture have attracted interest
from both academia and
industry~\cite{intel-disagg,fb-disagg,disstor-hotstor17,tsai2020disaggregating,legoOS}.
Modern public cloud providers such as Amazon and Microsoft Azure
Cloud~\cite{10.1145/2043556.2043571} have already adopted high-level compute
and storage disaggregation.

With these trends, more and more microservices or serverless computing-based
applications will be deployed and run on top of disaggregated architectures of
modern and next-generation datacenters~\cite{pemberton2019exploring}. In this
kind of deployments, we envision that next-generation microservices will be
more often offloaded and executed on remote computational devices or
disaggregated resources. In the meantime, these microservices need to be
migratable across different layers of heterogeneous hardware platforms, due to
the requirements of manageability, performance, load balancing, and so on. In
other words, we believe that offloadable and migratable microservices will
become a key computing paradigm on disaggregated datacenter architectures in
the future.

In this paper, we first present a brief survey on related work
(Section~\ref{sec:relatedwork}) to give a high-level overview of existing
studies in related research directions. Then, we outline a vision
(Section~\ref{sec:vision}) for demonstrating how and why microservices should
be offloaded and migratable on disaggregated datacenter architectures. The
associated challenges that distributed systems research may encounter are
discussed in Section~\ref{sec:challenges}. To address these fundamental
challenges, we propose a roadmap (Section~\ref{sec:roadmap}), mapping key
research challenges to bodies of work that have the potential to realize
high-performance, scalable, energy-efficient, and QoS-aware offloadable and
migratable microservice systems on disaggregated architectures.  The key
contributions of this paper are: 

\begin{itemize} 

    \item Envisioned a critical systems research direction of
designing and developing next-generation offloadable and migratable
microservices on emerging disaggregated datacenter architectures.
    
    \item Surveyed the recent related studies in the literature and the
community to demonstrate the importance and necessity of this research
direction.
    
    \item Outlined a vision and the associated fundamental challenges that
distributed systems and datacenter-scale computing research may encounter when
building next-generation offloadable and migratable microservices.
    
    \item Proposed a research roadmap to achieve our envisioned objectives in a
promising way. Within the roadmap, we identify potential techniques and methods
that can be leveraged.  

\end{itemize}

\section{A Brief Survey on Related Work}
\label{sec:relatedwork}

This section presents a brief survey on recent studies, which are closely
related to our research agenda.

\subsection{Microservice Benchmarking}

Research has been done on evaluating and benchmarking
microservices~\cite{DeathStarBench,ppbench,usuite,workload-microservice,benchmark-microservice,benchmarking-pattern}
but none of them discuss the pros and cons when making microservices
offloadable and migratable. DeathStarBench~\cite{DeathStarBench} contains
microservices that are representative of large end-to-end services and studies
their implications across the cloud system stack. It also shows that running
microservices on low power cores leads to slight performance degradation but
saves power. Thus, these cores could be utilized for microservices that are off
the critical path. \micro Suite~\cite{usuite} investigates the impact of
network and OS overheads on microservice median and tail latency. Ueda et
al.~\cite{workload-microservice} present the impact of language runtime and
hardware architecture on microservice performance. Zhou et
al.~\cite{benchmark-microservice} identify the gaps between existing benchmark
systems and industrial microservice systems and develop a new microservice
benchmarking system. ppbench~\cite{ppbench} proposes a benchmark to study the
performance impact of programming languages and networks on microservices.

\subsection{Disaggregated Datacenter Architecture}

Disaggregated datacenter architecture has many advantages like flexibility and
independent scaling of resources (compute, network, and/or storage),
fine-grained resource provisioning, and higher utilization. We believe this
architecture will continue to evolve and include new computational devices
which microservices should benefit from. A lot of work has been done in the
community to allow applications to reap the benefits of resource
disaggregation~\cite{asanovic2014hardware,shoal,legoOS,soNUMA,disagg-mem,disstor-hotstor17}.
LegoOS~\cite{legoOS} is a new OS model for hardware resource disaggregation
that splits traditional operating system functionalities into loosely-coupled
units. soNUMA~\cite{soNUMA} enables low latency and distributed in-memory
processing by eliminating data movement across network stack and PCIe
interface. Legtchenko et al.~\cite{disstor-hotstor17} discuss the concept of
disaggregated storage at rack-level and propose a novel storage fabric for
different disaggregation types for cloud storage.

\subsection{Near-Data Processing}

With the recent trends in disaggregated datacenter architectures and pushing
towards improving energy-efficiency, many applications are being re-designed to
move compute near to memory/storage. Near-Data Processing (NDP) has been
explored in the areas of databases~\cite{jafar,polarDB,dbMachines,queryDB},
computer vision~\cite{varifocal-storage}, machine-learning
inference~\cite{recssd}, big-data systems~\cite{biscuit}, and key-value
stores~\cite{Minerva,papyrusKV,nkv} with the vision to reduce data movement. We
believe microservices should also exploit this paradigm. Hu et
al.~\cite{varifocal-storage} present a storage system that adjusts the dataset
resolution within the storage device by utilizing idle SSD cores that are
otherwise used by flash translation layer and garbage collection inside an SSD.
E3~\cite{e3} is a microservice execution platform that offloads microservices
to low-power SmartNIC processors to improve energy-efficiency without much loss
in latency. E3 also migrates microservices to host to eliminate SmartNIC
overloading via a communication-aware service placement algorithm. Though some
of our motivations are similar to prior work, e.g., reducing energy consumption
on heterogeneous compute substrates, we not only want microservices to be
migratable to any available compute abstraction in the cluster but also to
benefit from NDP by leveraging computational devices in future datacenter
architectures without losing on the productivity of developers.

\subsection{Resource Virtualization \& Management}

Microservices are generally built on top of resource virtualization techniques
like containers and virtual machines (VMs). Apache OpenWhisk~\cite{openwhisk}
is an open-source, distributed serverless platform that executes functions in
response to events at any scale. OpenWhisk manages the infrastructure, servers,
and scaling using Docker containers. It supports many container frameworks and
programming languages in which one can write functional logic and run them in
response to events or from HTTP requests.

I/O virtualization introduces a major bottleneck for high-performance networks.
Remote Direct Memory Access (RDMA) virtualization is one way to achieve
migration of microservices in the new datacenter architecture across diverse
compute units. Current work in this space, namely,
SR-IOV~\cite{sriov,jie-ipdps17}, LITE~\cite{lite}, vSocket~\cite{vSocket},
HyV~\cite{HyV}, and FreeFlow~\cite{freeFlow} have heavy connection setup costs
and may not be suitable for container-based virtualized environments.
MasQ~\cite{MasQ} is a software-defined RDMA virtualization technique to allow
VMs in public clouds to realize virtual RoCE devices. MigrOS~\cite{MigrOS}
allows transparent live migration for RDMA-enabled containers.

\section{Vision}
\label{sec:vision}

\begin{figure}[t]
\includegraphics[width=0.49\textwidth]{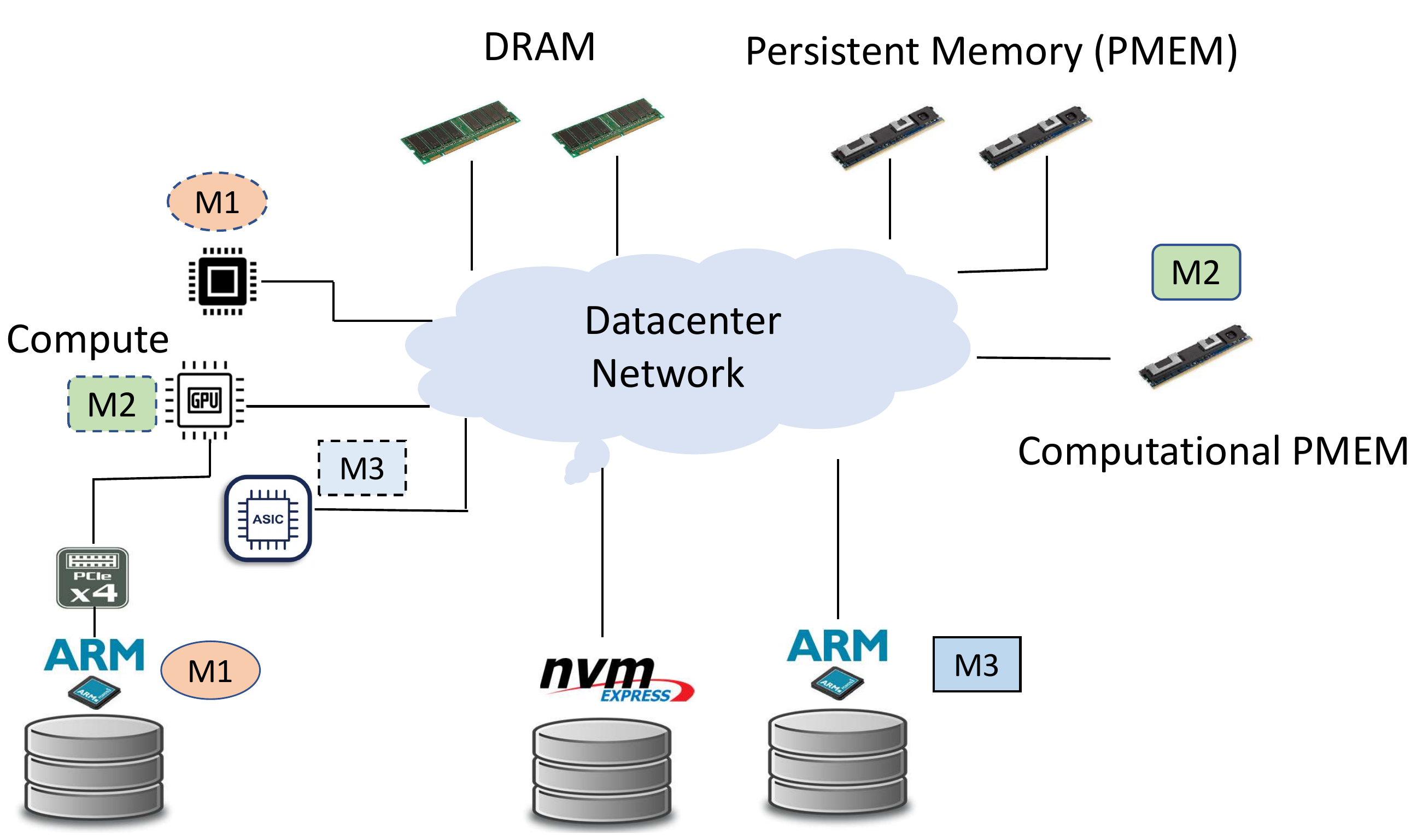}
\caption{Next-generation disaggregated datacenter architectures. M1, M2, and M3
are the microservices that were initially running on compute nodes and get
offloaded to computational devices to reduce data movement and achieve high
performance, energy-efficiency, and other benefits. } 
\label{fig:vision}
\end{figure}

As shown in Figure~\ref{fig:vision}, we envision the next-generation
disaggregated datacenter architectures to comprise of persistent
memory~\cite{fast-pmem,vldb-pmem} (PMEM), computational PMEM (CPM)~\cite{mcas},
NVMe-SSDs, computational SSDs (CSDs)~\cite{computational-ssd}, RDMA-enabled
networks, and remote memory~\cite{remote-regions,infiniswap}. Due to the
increasing opportunities of co-designing systems between hardware and software,
we want the microservices and serverless functions to be able to take advantage
of these new emerging hardware architectures in datacenters. Future
microservice systems should allow microservices and serverless applications to
leverage this new datacenter architecture by co-designing applications and
their underlying systems and runtime. This will improve their performance,
scalability,  energy-efficiency, QoS support while maintaining similar
productivity of the microservices developer.

There are a lot of low-level challenges in co-designing microservice
applications and frameworks with the emerging hardware. First, microservices
and serverless functions could be made up of different programming models and
languages adding to system heterogeneity and supporting them in diverse
hardware platforms becomes cumbersome. Second, as the computational storage
devices, like CSD and CPM, have limited compute and memory capacity (like
ARM-based SoCs) compared to x86 servers, utilizing their resources efficiently
without degrading their main functions is hard. Data-intensive microservices
that have less on-device memory consumption (like data scanning and filtering)
may be more appropriate to be offloaded to CSDs and CPMs.  Third, with the
introduction of new devices, application requirements of performance,
scalability, QoS, productivity, and energy-efficiency should be met. A
significant amount of energy and execution time consumed during an
application's lifetime goes into data movement between the compute and
memory~\cite{pim,nkv,polarDB}. Thus, the key insight here is to reduce this
data movement by pushing/offloading a certain amount of computing in
microservices or whole microservices near these computational devices.

Once microservices are offloaded on disaggregated resources, we will
immediately meet another requirement that is supporting migratable
microservices. This is because depending upon load distribution, data
distribution, and resource availability, microservices may need to be
rebalanced and migrated to other remote resources. As shown in
Figure~\ref{fig:vision}, microservices (e.g., M1, M2, and M3) were initially
running on local compute nodes. Later, they may be offloaded to some local or
remote computational devices (like CPM or CSD) due to data locality. At some
point, M1 or M3 may get migrated to CPM devices from ARM-based CSD platforms to
get better performance, energy efficiency, and load balancing. 

We envision that many application scenarios can easily get benefits from
offloadable and migratable microservices on disaggregated architectures. Taking
machine learning pipelines as an example, modern machine learning pipelines
including training and inference stages need not only a huge amount of
computing resources such as GPGPUs, TPUs, or ASICs but also data storage and
movement over networks and storage devices. A well-designed distributed system
such as Ray~\cite{ray} can significantly improve the performance,
manageability, and flexibility of machine learning pipelines. Machine learning
pipelines can be encapsulated to different microservices to get better
manageability and flexibility. In this paper, we envision that if
next-generation microservice management systems can smartly offload and migrate
microservices, then we can efficiently distribute training microservices,
pre-processing microservices, inferencing microservices, and post-processing
microservices in machine learning pipelines to the `best place' to be executed.
In this way, machine learning pipelines with our proposed microservices can
significantly reduce data movements, execution times, energy consumption, and
so on.

\section{Challenges}
\label{sec:challenges}

As indicated earlier, the availability of computational devices near storage
and memory, like CSDs and CPM, and next-generation datacenter platforms provide
novel opportunities to co-design microservices/serverless applications and
their underlying runtime to significantly improve energy-efficiency,
performance, scalability, QoS, and so on. However, much of the work to utilize
computational devices is still application-specific~\cite{nkv,polarDB,mcas} and
cannot be used by general-purpose microservices or serverless functions.

Data-intensive applications spend a significant amount of time and energy to
move data towards compute. But with the availability of compute units near the
memory and storage devices, certain tasks can be offloaded to these devices to
save energy consumption and improve the  performance of the application. Thus,
it becomes critical to co-design microservice applications to take advantage of
the next-generation datacenter hardware to reduce data movement and lessen
their carbon footprint. Based on these trends, we envision that hardware
evolution and cross-layer software co-design will continue to be the two most
important driving forces for the architectural evolution and
performance-boosting of future serverless computing systems. 

Based on these observations, we identify several fundamental research
challenges to achieve the goals of offloadable and migratable
microservices/serverless functions, which aim to fully utilize the provided
resources on modern and next-generation datacenter infrastructures.

\begin{enumerate}

\item How can we design effective and efficient runtime systems for
microservices and serverless functions that can take advantage of the novel
features of next-generation datacenter hardware?

\item How can we co-design with applications and microservice platforms to
achieve near-native performance while reducing energy consumption?

\item How to offload and migrate the entire microservice or a portion of it to
computational devices in an energy-efficient and QoS-aware way without
burdening any of resource-constrained devices? 

\item How to offload such tasks with minimal code changes to the applications
to keep the developer productivity high?

\item What kind of benefits can be achieved through these new designs? How to
benchmark these designs with microservice platforms and applications?

\end{enumerate}

\section{Research Roadmap}
\label{sec:roadmap}

\begin{figure*}[htbp] \begin{center} \mbox{ \subfloat
{\includegraphics[width=0.49\textwidth]{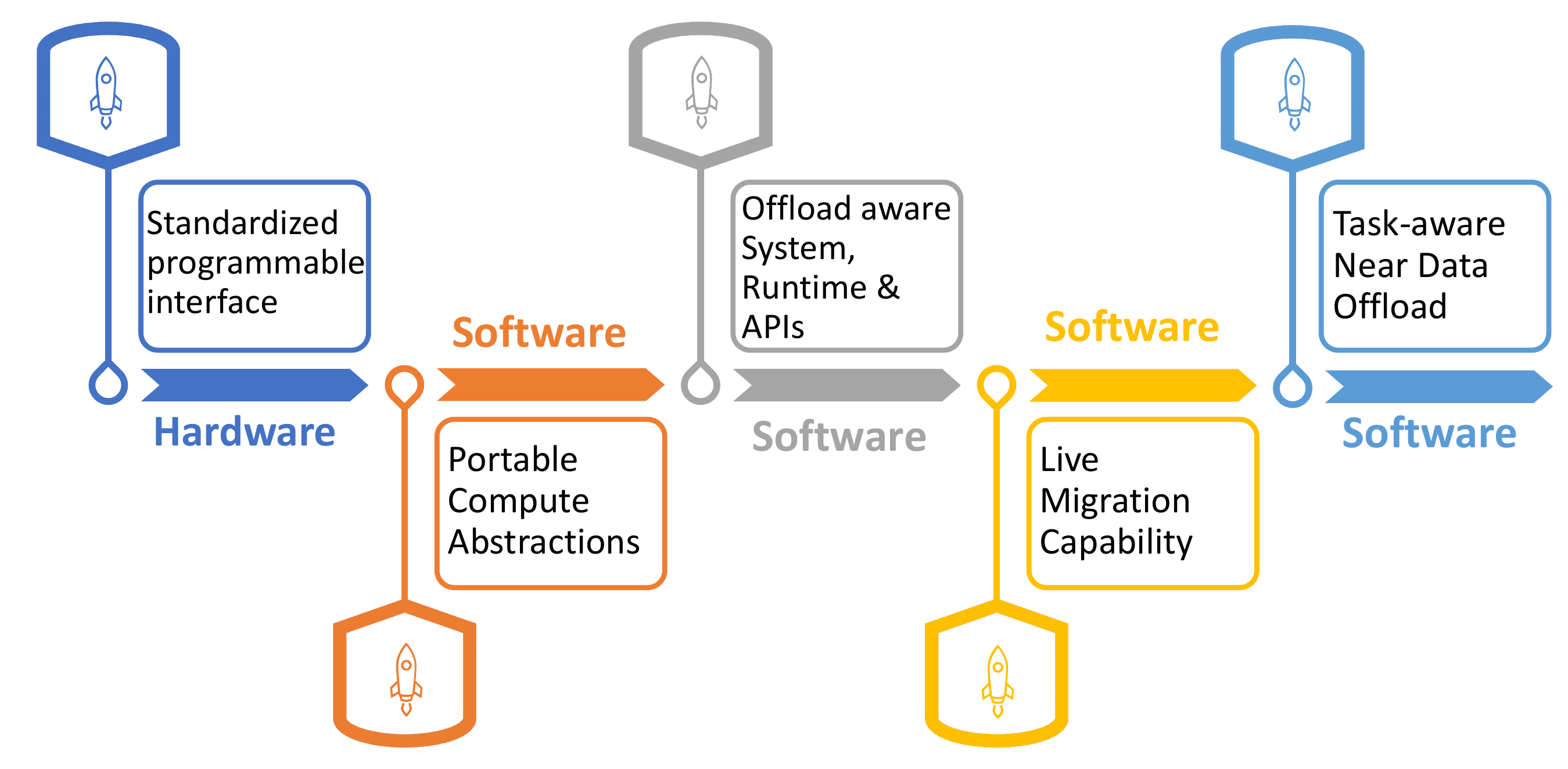} } \subfloat
{\includegraphics[width=0.49\textwidth]{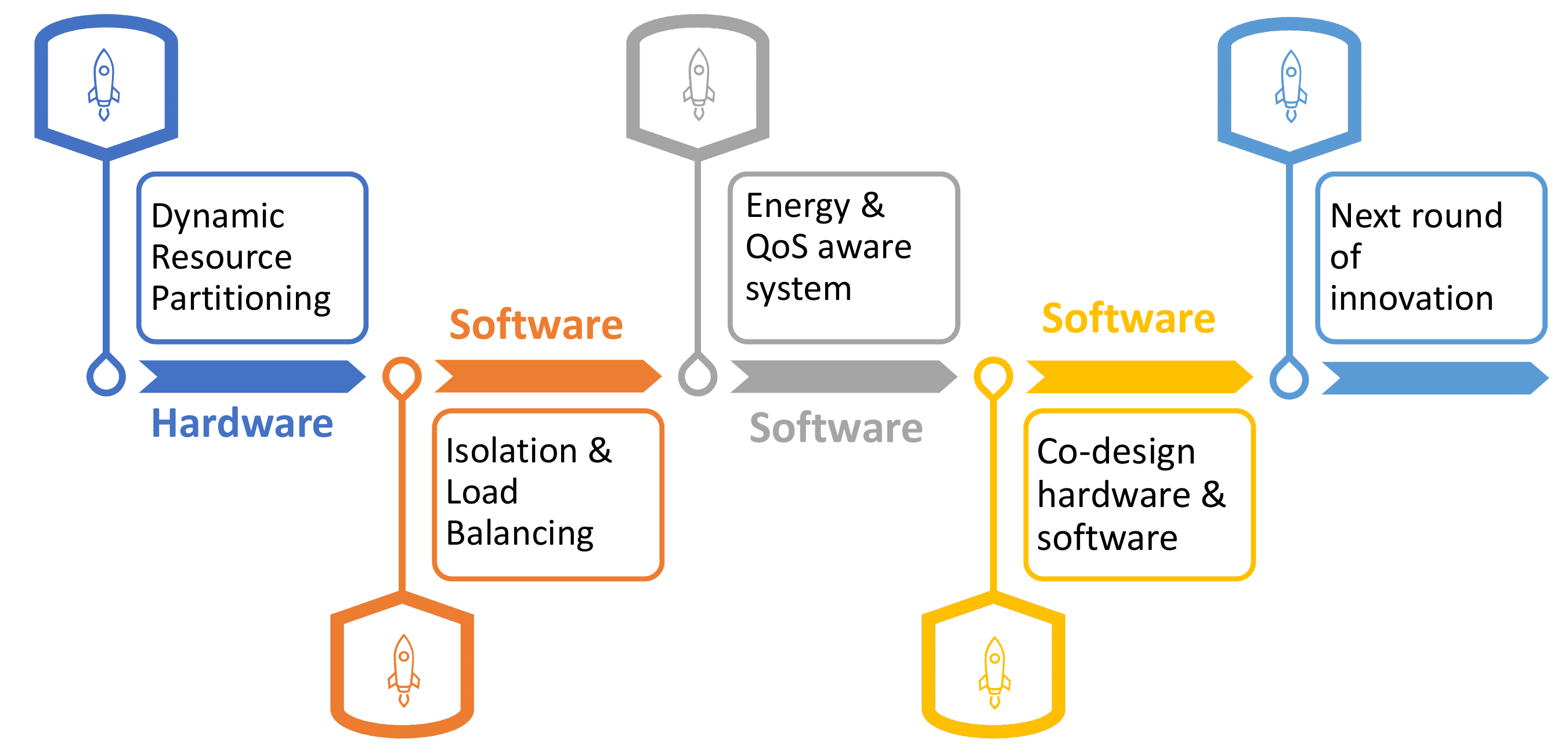}} }
\caption{Proposed Research Roadmap} \label{fig:roadmap} \end{center}
\end{figure*}

This section presents our proposed research roadmap as shown in
Figure~\ref{fig:roadmap}.

\subsection{Standardized Programmable Hardware Interface} 

First of all, it is a challenging task to offload an entire or a portion of a
microservice to heterogeneous compute units in the datacenter. This is because
we need a standardized, generic, and programmable interface from the underlying
hardware layer. Creating such an interface is challenging in a disaggregated
datacenter architecture as the heterogeneous computational devices could be
interconnected via the memory bus, intra-node, or inter-node interconnects
(like PCIe, CXL, Gen-Z, NVLink, RoCE, InfiniBand, etc.). Interacting with these
devices typically needs to change systems or applications a lot currently.
Hence, future research in this space should target offloading any functionality
across the datacenter in a way that has minimal impact on the productivity of
the developer and supports both built-in and resource-conservative user-defined
functions. Even though this is a difficult research task, we do see success
stories in the community. For instance, RDMA with verbs and NVMe protocols are
becoming standardized programmable interfaces for high-speed networks and
storage devices.  We envision that more standardized programmable interfaces
will become available in the future.

\subsection{Portable Compute Abstractions} 

When migrating/offloading microservices on various compute devices in the
datacenter, the ideal technique would be not to rewrite/recompile code for each
device. For instance, it may be a difficult problem to avoid the
cross-compilation issue when running microservices on different compute
devices. One possible solution is we could consider generating a number of
binaries and using the correct one on each compute device.

In addition, producing new compute abstractions would help us in reaching this
goal. Such abstractions should minimize the compute footprint in order to
reduce the compute movement cost across the datacenter and allow microservices
to transfer/share the output of offloaded compute. Thus, the community should
propose new abstractions which not only increase productivity but also maintain
portability and performance. So far, the most common ways to deploy
microservices are based on virtual machines or containers. To solve all the
challenges we listed earlier, it may be an interesting and open challenge to
rethink what could be the best way to design portable compute abstractions for
microservices.

\subsection{Offload-aware Systems, Runtimes, and APIs} 

Microservices could run numerous kinds of workloads, which can be both stateful
and stateless workloads.  We foresee different approaches should be adopted to
handle stateful and stateless microservices.  In order to fully support various
workloads and execute offloaded functions on different kinds of computational
devices, we need offload-aware systems, runtimes, and APIs.  Developing such
kind of systems to fully leverage diverse compute units in an
application-agnostic manner and at the same time add minimal overhead is an
interesting research direction.  For example, an offload-ware communication and
I/O runtime should be able to automatically and transparently select the best
protocols for microservice applications based on the capabilities of the
devices where the microservices are offloaded to.

\subsection{Live Migration Capability} 

Microservice migration, especially for stateful microservices, is important to
avoid overloading and achieve good utilization across computational devices in
a datacenter. Migration needs not to be just from one host/server to another.
It could be, for example, from a local computational storage device to a remote
computational persistent memory device or the other way around. We need to
develop new mechanisms that enable fast live migration of microservices using
different network protocols and interconnects while preserving performance.
Another important aspect of enabling live migration would be to figure out
which heterogeneous compute unit has the resources to support the migrating
microservice without violating the QoS support.

\subsection{Task-Aware Near Data Offload} 

For maximum utilization of available compute resources and to reduce data
movement and energy consumption, task and capability aware offloading of
microservices is necessary. A good offload strategy should decide where to
offload and also take into consideration offloading costs vs. rewards.  Future
research in task-aware offloading should explore designing new metrics and
models which could characterize and predict when and on which computational
device it would be useful to offload microservices in disaggregated
datacenters. The potential modeling and decision-making approaches can be built
with some artificial intelligence techniques, such as reinforcement learning. 

\subsection{Dynamic Resource Partitioning} 

Due to the heterogeneous nature of compute, memory, and workloads on a
computational device in a datacenter, dynamic resource partitioning for
offloaded functions is a challenge. One must ensure that running microservices
or functions on CSDs/CPMs should not impact the performance of their main task
-- I/O processing. To achieve this, hardware vendors may need to provide more
resources or mechanisms on hardware devices to help systems designs. For
example, dynamic I/O queues or resource pools can be provided from the hardware
layer to achieve better designs.

\subsection{Isolation and Load Balancing} 

As these computational devices will run multiple microservices/offloaded
functions, there should be strong security and isolation guarantees among them
as present on a host server.  Some possible technology examples include
hardware-based data encryption, namespaces, protection domains, etc.

Also, the load balancer which would analyze the load of all or nearby devices
should make offloading decisions based on data locality and device load.  To
achieve scalable load balancing, both centralized and decentralized load
balancing algorithms should be designed. 

\subsection{Energy and QoS Aware Systems} 

All microservices have to meet their Service Level Objectives (SLOs), so
offloading of microservices, especially compute-intensive in nature, should
occur if running them on computational devices does not violate their QoS
guarantees. Another important factor to consider while executing microservices
is to respect the pre-defined energy/power budget. Hence, the challenge would
be to balance the performance-energy trade-off based on varying application and
user needs. These goals can be achieved through accurate modeling and
enforcement in microservice systems.  Artificial intelligence techniques may
again play a very important role in this research direction.

\subsection{Co-design Hardware \& Software} 

Multiple layers of software frameworks, programming model semantics, and newer
architectures have rendered the design process to be complicated and isolated.
Thus, a holistic cross-layer approach is critical to tackling the challenges at
these layers. The community needs to develop runtimes or systems that could
provide efficient virtualization mechanisms, live migration capability, and
support all the other research challenges outlined above. Then microservices
and serverless functions could be co-designed to coordinate and interact with
the underlying runtimes and reap all the benefits of next-generation datacenter
infrastructures. \\

We envision the next round of potential innovation to happen in another 5-10
years after materializing the research roadmap discussed here. This research
roadmap needs a lot of research work and innovations from the community to
achieve these goals. 

\section{Conclusion}

This vision paper emphasized an important systems research direction on
offloadable and migratable microservices on disaggregated datacenter
architectures. We performed a systematic survey on related studies, presented
our vision and associated research challenges, and proposed a research roadmap
to achieve the envisioned objectives. In the future, we plan to build a new
microservice or serverless computing system, which can efficiently support the
features outlined in our research agenda. We also wish our vision paper can
bring more attention from the community to collaborate and participate in this
research direction.

\section*{Acknowledgement}
\label{sec:ack}
We would like to thank Stanko Novakovic and anonymous reviewers for their valuable feedback. This work was supported in part by NSF research grant CCF \#1822987.

\bibliographystyle{plain}

\end{document}